\documentclass[useAMS,usenatbib]{mn2e}

\title{Whistler Wave Cascades in Solar Wind Plasma}
\author[Dastgeer Shaikh]
  {Dastgeer ~Shaikh,$^1$\thanks{email: dastgeer.shaikh@uah.edu} \\
  $^1$  Department of Physics and Center for Space Plasma and Aeronomic Research (CSPAR)\\
The University of Alabama in Huntsville,
Huntsville. Alabama, 35899}
\date{Revised on Feb 20, 2009}

\pagerange{\pageref{firstpage}--\pageref{lastpage}} \pubyear{2009}

\def\LaTeX{L\kern-.36em\raise.3ex\hbox{a}\kern-.15em
    T\kern-.1667em\lower.7ex\hbox{E}\kern-.125emX}

\newcommand{\be}{\begin{equation}}
\newcommand{\ee}{\end{equation}}

\newcommand{\eq}[1]{Eq. (\ref{#1})}

 \newcommand{\eqa}{\begin{eqnarray}}
\newcommand{\eeq}{\end{eqnarray}}

\begin{document}

\label{firstpage}

\maketitle

\begin{abstract}
Nonlinear three dimensional, time dependent, fluid simulations of
whistler wave turbulence are performed to investigate role of whistler
waves in solar wind plasma turbulence in which characteristic
turbulent fluctuations are characterized typically by the frequency
and length scales that are respectively bigger than ion gyro frequency
and smaller than ion gyro radius.  The electron inertial length is an
intrinsic length scale in whistler wave turbulence that
distinguishably divides the high frequency solar wind turbulent
spectra into scales smaller and bigger than the electron inertial
length. Our simulations find that the dispersive whistler modes evolve
entirely differently in the two regimes. While the dispersive whistler
wave effects are stronger in the large scale regime, they do not
influence the spectral cascades which are describable by a
Kolmogorov-like $k^{-7/3}$ spectrum.  By contrast, the small scale
turbulent fluctuations exhibit a Navier-Stokes like evolution where
characteristic turbulent eddies exhibit a typical $k^{-5/3}$
hydrodynamic turbulent spectrum. By virtue of equipartition between
the wave velocity and magnetic fields, we quantify the role of
whistler waves in the solar wind plasma fluctuations.

\end{abstract}

\begin{keywords}
 (magnetohydrodynamics) MHD, (Sun:) solar wind, Sun: magnetic fields, ISM: magnetic fields
\end{keywords}

\section{Introduction}
The solar wind is an excellent in-situ laboratory for investigating
nonlinear and turbulent processes in a magnetized plasma fluid since
it comprises a multitude of spatial and temporal length-scales
associated with an admixture of waves, fluctuations, structures and
nonlinear turbulent interactions. The in-situ spacecraft measurements
(Matthaeus \& Brown 1988, Goldstein et al 1995) reveal that the solar
wind fluctuations, extending over several orders of magnitude in
frequency and wavenumber, describe the power spectral density (PSD)
spectrum that can be divided into three distinct regions (Goldstein et
al 1995, Leamon et al 1999). The frequencies, for instance, smaller
than 10-5 Hz lead to a PSD that has a spectral slope of -1 . This
follows the region that extends from 10-5 Hz to or less than
ion/proton gyrofrequency where the spectral slope exhibits an index of
-3/2 or -5/3. The latter, a somewhat controversial issue, is
characterized essentially by fully developed turbulence and can be
followed from the usual magnetohydrodynamic (MHD) description. The
spacecraft observations have further revealed that length scales
beyond the MHD regime, that are smaller than ion gyro radius ($k\rho_i
\gg 1$) and temporal scales bigger than ion cyclotron frequency
$\omega> \omega_{ci}=eB_0/m_ec$, (where $k, \rho_i, \omega_{ci}, e,
B_0, m_e, c$ are respectively characteristic mode, ion gyroradius, ion
cyclotron frequency, electronic charge, mean magnetic field, mass of
electron and speed of light) exhibit a spectral break where the
inertial range slope of the solar wind turbulent fluctuations varies
between -2 and -5 (Smith et al 1990, Goldstein et al 1994, Leamon et
al 1999).  Notably, the dynamics of the length-scales in this region
cannot be described by the usual MHD models that possess
characteristic frequencies smaller than ion gyro frequencies.  At 1
AU, the ion inertial length scales are smaller than ion gyro radii in
the solar wind \cite{Goldstein1995}. The latter, associated with
plasma motion due to finite Larmor radii, can readily be resolved in
the usual MHD models by introducing Hall terms to accommodate the ion
gyro scales up to scales as small as ion inertial length scales.  The
higher time resolution databases identifying the spectral break
indicate that Alfvenic MHD cascades (Smith et al 1990, Goldstein et al
1994, Leamon et al 1999) are terminated near the spectral break. The
characteristic modes in this region are observed to evolve typically
on timescales involving the dispersive kinetic Alfvenic
fluctuations. The onset of the second or the kinetic Alfven inertial
range is still elusive to our understanding of the solar wind
turbulence and many other nonlinear interactions. Specifically, the
mechanism leading to the spectral break has been thought to be either
mediated by the kinetic Alfven waves (KAWs) (Hasegawa 1976), or by
electromagnetic ion-cyclotron-Alfven (EMICA) waves \cite{gary,yoon},
or by a class of fluctuations that can be dealt within the framework
of the HMHD plasma model (Alexandrova et al 2007, 2008; Shaikh \&
Shukla 2008, 2008a). Stawicki et al (2005) argue that Alfv\'en
fluctuations are suppressed by proton cyclotron damping at
intermediate wavenumbers so the observed power spectra are likely to
consist of weakly damped magnetosonic and/or whistler waves which are
dispersive unlike Alfv\'en waves. Moreover, turbulent fluctuations
corresponding to the high frequency and $k\rho_i \gg 1$ regime lead to
a decoupling of electron motion from that of ion such that the latter
becomes unmagnetized and can be treated as an immobile neutralizing
background fluid.  While whistler waves typically survive in the
higher frequency (and the corresponding smaller length scales) part of
the solar wind plasma spectrum, their role in influencing the inertial
range turbulent spectral cascades is still debated
\cite{biskamp,dastgeer03,dastgeer05,dastgeer08,dastgeer08a}.  The
Kolmogorov like dimensional arguments indicate that propagation of
whistlers in the presence of a mean or an external constant magnetic
field may change the spectral index of the inertial range turbulent
fluctuations from $k^{-7/3}$ to $k^{-2}$ \cite{biskamp}. By contrast,
the numerical simulations \cite{biskamp,dastgeer03,dastgeer05} suggest
that whistler waves do not influence the spectral migration of
turbulent energy in the inertial range despite strong wave activity
and that the turbulent spectra corresponding to the electron fluid
fluctuations in whistler wave turbulence continue to exhibit a
Kolmogorov-like $k^{-7/3}$ spectrum. What is not clear from these work
is the quatitative role of whistler and the corresponding mode
coupling interactions that mediate the inertial range turbulent
spectra. Furthermore, the whistler wave turbulence described in the
Refs. \cite{biskamp,dastgeer03,dastgeer05} focus on purely two
dimensional interactions, ignoring thus the variations in the third
dimension. It is therefore unclear whether the nonlinear whistler mode
coupling interactions in three dimensions modify the energy cascades
in the inertial range turbulence.  Intrigued largely by these issues,
the primary goal of this paper is to investigate the nonlinear
interaction amongst whistler waves and turbulent fluctuations, based
on nonlinear fluid simulations, in $\omega> \omega_{ci}$ regime where
correlation length scales of turbulence are comparable to the electron
inertial length scales.  Understanding the role of whistler waves in
the turbulent cascades is crucial to many other space plasma processes
since whistler waves, in addition to solar wind turbulence, are
instrumental in governing nonlinear processes in numerous other plasma
systems that range from solar wind
\cite{krafft,saito,Stawicki,gary,ng,Vocks,salem,Bhattacharjee1998},
magnetic reconnection in the Earth's magnetosphere \cite{Wei2007} to
interstellar medium \cite{burman} and astrophysical plasmas
\cite{roth} where characteristic fluctuations can typically be of
several astronomical units. These are only a few of the numerous other
studies. For more literature, the readers can refer to the simulation
work by Biskamp \cite{biskamp} and others including Shukla (1978),
Shukla et al (2001), Cho \& Lazarian (2004), Galtier (2008), Urrutia
et al (2008), Saito et al (2008), Bengt \& Shukla (2008), Shaikh
(2009) and numerous references therein.

In this paper, I focus on understanding the nonlinear turbulent
cascades mediated by whistler waves in a fully three dimensional
geometry. Our objective is to investigate the role of whistlers in
establishing the turbulent equipartition amongst the modes that are
responsible for the nonlinear mode coupling interactions which
critically determine the inertial range power spectra. Remarkably, we
find that despite the equipartition processes mediated by whistler
modes for which the wave activity is strong, the inerital range
spectra continue to exhibit a Kolmogorov-like spectrum where whistler
wave effects are {\em unimportant}.  We begin in Section 2 by
describing the underlying whistler wave turbulence model and it's
linear properties. Section 3 describes nonlinear simulation results of
inertial range turbulent spectra. In section 4, we discuss the
theoretical arguments corresponding to the whistler wave turbulent
spectra that correspond to the characteristic length scales smaller as
well as bigger than electron inertial length ($d_e$).  The process of
turbulent equipartition between the magnetic and velocity field,
quantifying the whistler wave effects, is also desribed in this
section. Finally, section 5 summarizes our results.

\section{Whistler Wave Turbulence Model}
Whistler modes are excited in the solar wind plasma when the
characteristic plasma fluctuations propagate along a mean or background
magnetic field with frequency $\omega>\omega_{ci}$ and the length
scales are $c/\omega_{pi} < \ell < c/\omega_{pe}$, where $\omega_{pi},
\omega_{pe}$ are the plasma ion and electron frequencies.  The
electron dynamics plays a critical role in determining the nonlinear
interactions while the  ions merely provide a stationary
neutralizing background against fast moving electrons and behave as
scattering centers. The whistler wave turbulence can be described by
the electron magnetohydrodynamics (EMHD) model of plasma \citep{model}
that deals with the single fluid description of quasi neutral plasma.
The EMHD model has been discussed in considerable detail in earlier
work \citep{model,biskamp,dastgeer00a,dastgeer00b,dastgeer03,dastgeer05}.  In
whistler modes, the currents carried by the electron fluid are
important, and we therefore write down only those equations which are
pertinent to electron motion. These are electron fluid momentum, electric field, currents, and
electron continuity equations,
\be
\label{elec}
m_e n \frac{\partial {\bf V}_e}{\partial t} + {\bf V}_e \cdot \nabla {\bf V}_e
=-en {\bf E} -  \frac{ne}{c} {\bf V}_e \times {\bf B} - \nabla P
-\mu m_e n {\bf V}_e, 
\ee
\be
{\bf E} = -\nabla \phi - \frac{1}{c} \frac{\partial {\bf A}}{\partial t},
\ee
\be
\label{ampere}
\nabla \times {\bf B} = \frac{4\pi}{c} {\bf J} +  
\frac{1}{c} \frac{\partial {\bf E}}{\partial t},
\ee
\be
\frac{\partial n}{\partial t} + \nabla \cdot (n {\bf V}_e) = 0.
\ee
The remaining equations are ${\bf B} = \nabla \times {\bf A}, {\bf J}
= -en{\bf V}_e, \nabla \cdot {\bf B} =0$. Here $m_e, n, {\bf V}_e$ are
the electron mass, density and fluid velocity respectively. ${\bf E},
{\bf B}$ respectively represent electric and magnetic fields and $\phi
, {\bf A}$ are electrostatic and electromagnetic potentials. The
remaining variables and constants are, the pressure $P$, the
collisional dissipation $\mu$, the current due to electrons flow ${\bf
J}$, and the velocity of light $c$.  The displacement current in
Ampere's law \eq{ampere} is ignored, and the density is considered as
constant throughout the analysis.  The electron continuity equation
can therefore be represented by a divergence-less electron fluid
velocity $\nabla \cdot {\bf V}_e = 0$. The electron fluid velocity can
then be associated with the rotational magnetic field through
\be
\label{velocity}
{\bf V}_e = - \frac{c}{4\pi n e}\nabla \times{\bf B}.
\ee
On taking the  curl of \eq{elec} and, after slight rearrangement
of the terms, we obtain

\be
\label{PP}
\frac{\partial {\bf P}}{\partial t} - {\bf V}_e  \times ( \nabla \times {\bf P})
+\nabla \xi = - \mu m_e {\bf V}_e
\ee
where
\[
{\bf P} = m_e{\bf V}_e - \frac{e{\bf A}}{c} ~~~~~~~{\rm and}~~~
\xi = \frac{1}{2} m_e {\bf V}_e\cdot {\bf V}_e + \frac{P}{n} - e\phi.
\]
Here ${\bf P}$ is  generalized electron momenta. The curl of \eq{PP} eliminates the gradient of
the scalar quantity (the third term from the left in the lhs) and  yields
\be
\frac{\partial {\bf \Omega}}{\partial t} + 
\nabla \times ({\bf V}_e  \times {\bf \Omega} ) = - \mu m_e \nabla \times{\bf V}_e,
\ee
where
\[
{\bf \Omega} = \nabla \times{\bf P} = d_e^2 \nabla^2 {\bf B} - {\bf B}.
\]
It can be seen from Eq. (6) that in the ideal whistler mode turbulence
(i.e. neglecting the term associated with the damping $\mu$), the Curl
of generalized electron momenta is frozen in the electron fluid
velocity.  This feature is strikingly similar to Alfv\'enic turbulence
where the magnetic field is frozen in the ideal two fluid plasma
\cite{biskamp03}.  On substituting ${\bf \Omega}$ into the above
equation and using appropriate vector identities, we obtain the
three-dimensional equation of EMHD describing the evolution of the
magnetic field fluctuations in whistler wave,
\eqa
\label{emhd3}
\frac{\partial }{\partial t}({\bf B}-d_e^2 \nabla^2 {\bf B} ) + {\bf
V}_e\cdot \nabla ({\bf B}-d_e^2 \nabla^2 {\bf B} )- \\ \nonumber ({\bf B}-d_e^2
\nabla^2 {\bf B} ) \cdot \nabla{\bf V}_e 
 = \mu d_e^2 \nabla^2 {\bf B}.
\eeq 
The length scales in \eq{emhd3} are normalized by the electron skin
depth $d_e = c/\omega_{pe}$ i.e. the electron inertial length scale,
the magnetic field by a typical amplitude ${B}_0$, and time by
the corresponding electron gyro-frequency.  In \eq{emhd3}, the
diffusion operator on the right hand side is raised to $2n$. Here $n$
is an integer and can take $n=1,2,3, \cdots$.  The case $n=1$ stands
for normal diffusion, while $n=2,3, \cdots$ corresponds to hyper- and
other higher order diffusion terms.

The linearization of \eq{emhd3} about a constant magnetic field ${\bf
  B}=B_0\hat{z}+\tilde{\bf B}$, where $B_0$ and $\tilde{\bf B}$ are
respectively constant and wave magnetic fields, yields the following
equation,
\be
\omega_k (1+d_e^2k^2) \tilde{\bf B} + \frac{CB_0}{4\pi ne} ik_{\parallel} {\bf k} \times \tilde{\bf B}=0.
\ee
On eliminating the wave perturbed magnetic field from the above relation, one obtains
the following dispersion relation,
\be
\label{disp}
\omega_k = \omega_{c_0}\frac{d_e^2 k_yk}{1+d_e^2k^2},
\ee
where $ \omega_{c_0}=eB_0/mc$ and $k^2=k_x^2+k_y^2$.  
The use of \eq{disp} in Eq. (9) leads to the following relation
between the wave magnetic field and the velocity field,
\be
\tilde{\bf B}= \pm \frac{i}{k}  {\bf k} \times \tilde{\bf B}
\ee
The rhs of Eq. (11), in combination with \eq{velocity}, corresponds
essentially to the whister wave perturbed velocity field. This
equation indicates that whistler waves are transverse and are produced
by rotational magnetic field that leads essentially to the velocity
field fluctuations. On replacing the rhs in Eq. (11) with the
perturbed velocity field, it can be shown that the whistler modes obey
equipartition between the magnetic and velocity field components as
$k^2|B|^2 \simeq |V_e|^2$. The whistler wave activity can thus be
quantified by how closely the characteristic modes obey the turbulent
equipartition relation.  In section 4, we will investigate using nonlinear 3D
fluid simulations the equipartition mediated by whistler waves in the
inertial range turbulent spectra to signify the role of whistlers in
the solar wind plasma.

It becomes evident from \eq{emhd3} that there exists an intrinsic
length scale $d_e$, the electron inertial skin depth, which divides
the entire turbulent spectrum into two regions; namely short scale
($kd_e>1$) and long scale ($kd_e<1$) regimes.  In the regime $kd_e<1$,
the linear frequency of whistlers is $\omega_k \sim k_y k$ and the
waves are dispersive.  Conversely, dispersion is weak in the other
regime $kd_e>1$ since $\omega_k \sim k_y/ k$ and hence the whistler
wave packets interact more like the eddies of hydrodynamical fluids.
The equation of EMHD [\eq{emhd3}] is also exactly integrable, yielding
the total energy integral;
\[
E= \frac{1}{2}\int (1+d_e^2 \nabla^2) |{\bf B}|^2 d^3{\bf v},\]
and generalized helicity. Here $ d^3{\bf v}$ is a 3D volume element.
In the presence of dissipation ($\mu$) the total energy decays
eventually with time since 
\be \frac{\partial E}{\partial t} = - \mu d_e^2
\int d^3{\bf v} \left[ (\nabla^2 B)^2 \right].
\label{energy}
\ee 
Hence the inclusion of dissipation will damp the smaller scale fluctuations in the
whistler wave turbulence. The damping of the smaller dissipative scales are not
expected to influence the inertial range turbulent cascades.

\section{Simulation results}

\begin{figure}
\vspace{302pt}
\begin{center}
\end{center}
\caption{ 3D simulation of whistler wave turbulence in the $kd_e<1$
  regime exhibits a Kolmogorov-like inertial range power spectrum
  close to $k^{-7/3}$. The simulation parameters are: Box size is $L_x
  \times L_y \times L_z=2\pi \times 2\pi \times 2\pi$, numerical
  resolution is $N_x \times N_y \times N_z = 200 \times 200 \times
  200$, electron skin depth is $d_e=0.015$, magnitude of constant
  magnetic field is $B_0=0.5$. The characteristic large scales in this
  regime possess strong dispersion and wave activity that can be
  quantified from the turbulent equipartition between the velocity and
  magnetic fields. }
\label{fig1}
\end{figure}

\begin{figure}
\vspace{302pt}
\begin{center}
\end{center}
\caption{The small scales magnetic field fluctuations in the $kd_e>1$
  regime depicts a Kolmogorov-like $k^{-5/3}$ spectrum which is a
  characteristic of hydrodynamic fluid. The simulation parameters are
  same as those used in Fig 1, except $d_e=0.15$. Our simulations show that
  the small scale fluctuations evolve as non magnetized eddies of
  hydrodynamic fluid where whistler waves do not influence the energy
  cascades.}
\label{fig2}
\end{figure}

Turbulent interactions mediated by the coupling of whistler waves and
inertial range fluctuations are studied in three dimensions (3D) based
on a nonlinear 3D whistler wave turbulence code that we have developed
at Center for Space Plasma and Aeronomic Research (CSPAR), the
University of Alabama in Huntsville (UAH). Our code numerically
integrates \eq{emhd3}.  The spatial descritization employs a
pseudospectral algorithm \cite{scheme,dastgeer06,dastgeer07} based on
a Fourier harmonic expansion of the bases for physical variables
(i.e. the magnetic field, velocity), whereas the temporal integration
uses a Runge Kutta (RK) 4th order method.  The boundary conditions are
periodic along the $x,y$ and $z$ directions in the local rectangular
region of the solar wind plasma.

The turbulent fluctuations are initialized by using a uniform
isotropic random spectral distribution of Fourier modes concentrated
in a smaller band of lower wavenumbers ($k<0.1~k_{max}$). While
spectral amplitudes of the fluctuations are random for each Fourier
coefficient, it follows a certain initial spectral distribution
proportional to $k^{-\alpha}$, where $\alpha$ is an initial spectral
index.  The spectral distribution set up in this manner initializes
random scale turbulent fluctuations.  We note that a constant magnetic
field is included along the $z$ direction (i.e. ${\bf B}_0 =B_0 \hat
{\bf z}$) to accommodate the large scale (or the background solar
wind) magnetic field. The size of the 3D computational domain is
$(2\pi)^3$ with the spectral resolution $256^3$. In this paper, we
present the results of freely decaying whistler wave turbulence and
focus primarily on understanding the inertial range cascades in both
the $kd_e<1$ and $kd_e>1$ regimes. In principle, turbulence can be
driven.  The driven whistler turbulence is nonetheless beyond the
scope of this paper.

Electron whistler fluid fluctuations, in the presence of a constant
background magnetic field, evolve by virtue of nonlinear interactions
in which larger eddies transfer their energy to smaller ones through a
forward cascade. According to \cite{kol}, the cascades of spectral
energy occur purely amongst the neighboring Fourier modes (i.e. local
interaction) until the energy in the smallest turbulent eddies is
finally dissipated gradually due to the finite dissipation. This leads
to a damping of small scale motions.  By contrast, the large-scales
and the inertial range turbulent fluctuations remain unaffected by
direct dissipation of the smaller scales. Since there is no mechanism
that drives turbulence at the larger scales in our model, the
large-scale energy simply migrates towards the smaller scales by
virtue of nonlinear cascades in the inertial range and is dissipated
at the smallest turbulent length-scales.  The spectral transfer of
turbulent energy in the neighboring Fourier modes in whistler wave
turbulence follows a Kolmogorov phenomenology \citep{kol, iros, krai}
that leads to Kolmogorov-like energy spectra. We find from our 3D
simulations that whistler wave turbulence in the $kd_e<1$ and $kd_e>1$
regimes exhibits respectively $k^{-7/3}$ (see Fig 1) and $k^{-5/3}$
(see Fig 2) spectra. The inertial range turbulent spectra obtained
from our 3D simulations are further consistent with 2D work
\cite{biskamp,dastgeer00a,dastgeer00b}.  Interestingly, it is evident
from the whistler wave dispersion relation that the wave effects
dominate in the large scale, i.e. $kd_e<1$, regime where the inertial
range turbulent spectrum depictes a Kolmogorov-like $k^{-7/3}$
spectrum. On the other hand, turbulent fluctuations in the smaller
scale ($kd_e>1$) regime behave like non magnetic eddies of
hydrodynamic fluid and yield a $k^{-5/3}$ spectrum.  The wave effect
is weak, or negligibly small, in the latter. Hence the nonlinear
cascades are determined essentially by the hydrodynamic like
interactions. The observed whistler wave turbulence spectra in the
$kd_e<1$ and $kd_e>1$ regimes (Figs 1 \& 2) can be followed from the
Kolmogorov-like arguments \citep{kol, iros, krai} that describe the
inertial range spectral cascades. We elaborate on these arguments to
explain our simulation results of Fig. (1) \& (2) in the following
section.

\section{Energy spectra in whistler wave turbulence}

The exact spectral indices corresponding to the whistler wave
turbulent spectra, described by the ideal electron magnetohydrodynamic
invariant, can be understood from the Kolmogorov's dimensional
arguments \cite{kol, iros, krai}. The electron skin depth, $d_e$, in
EMHD turbulence intrinsically divides the entire Fourier spectrum into
regions for which length scales are either larger or smaller than
$d_e$. We derive the spectral indices for both regions of the
turbulent spectrum.

In the underlying whistler wave model of magnetized plasma turbulence,
the inertial range eddy velocity is characterized typically by $v_e
\sim \nabla \times{\bf B}$. Thus the typical velocity of the magnetic
field eddy $B_{\ell}$ with a scale size $\ell$ can be represented by
$v_e \simeq B_{\ell}/\ell$. The eddy turn-over time is then given by
\[\tau \sim \frac{\ell}{v_e} \sim \frac{\ell^2}{B_{\ell}}.\]
This is the time scale that predominantly leads to the nonlinear
spectral transfer of energy in fully developed whistler wave
turbulence. While the inertial range nonlinear cascades are determined
essentially by the eddy turn over or spectral transfer time scale, it
is not clear whether the characteristic length scales bigger than
$d_e$, where whistler wave propagation dominate, are influenced by
whistler interaction time scales. We will comment on this issue in the
following alongwith the inertial range spectra in both the $kd_e<1$
and $kd_e>1$ regimes.

\begin{figure}
\begin{center}
\vspace{302pt}
\end{center}
\caption{Turbulent equipartion between the velocity and magnetic
  fields is observed in our 3D simulations. The equipartition is
  measured for the entire turbulent spectrum at each time step by the
  relation $E_{equi} \simeq \sum_k (|v_e|^2 - k^2|B_k|^2)$. When the
  characteristic turbulent modes evolve towards equipartition, the
  relationship $|v_{e}({\bf k},t)|^2 \simeq k^2|B({\bf k},t)|^2$ is
  obeyed. Consequently, $E_{equi} \rightarrow 10^{-7}$. This number is
  small enough to establish a nearly perfect equipartition between the
  velocity and magnetic field associated with the whistler waves.}
\label{fig3}
\end{figure}

\subsection{$kd_e < 1$: Whistler wave  regime}
In the regime where characteristic length scales are bigger than the
electron skin depth ($kd_e < 1$), the inertial range whistler
turbulent energy is dominated by the large scale fluctuations. The
total energy corresponding to the turbulent fluctuations in this
regime is then given as,
\[ E \sim |{\bf B}|^2 \sim B_{\ell}^2 \sim v_e^2 \ell^2. \]
The $B_{\ell}$ represent magnetic field associated with the magnetic
field eddy of length $\ell$. The second similarity follows from the
assumption of an equipartition of energy in the magnetic and velocity
field components of whistler waves. The process of equipartition
origintes from the correlation between the velocity and magnetic field
fluctuations $v_e \sim {\bf k} \times{\bf B}$, where ${\bf
  k}=k_x\hat{x}+k_y \hat{y}+ k_z \hat{z}$ is a three dimensional wave
vector. The latter is further consistent with the electron flow speed,
i.e. Eq (6) in combination with the wave perturbed magnetic field
Eq. (11), that is used to derive the dynamical equation of whistler wave
turbulence, i.e. \eq{emhd3}.  This velocity-magnetic field correlation
essentially produces the velocity field fluctuations that are normal
to the magnetic field in a whistler wave packet. Consequently, the
energy associated with the velocity and magnetic field for each
characteristic turbulent mode evolves toward a relationship that satisfies $v_e^2
\simeq k^2 B^2$. To quantify our arguments, we follow the
evolution of turbulent equipartion in our simulations by computing the
following quantity, \be E_{equi}(t) \simeq \sum_k (|v_e(k,t)|^2 -
k^2|B(k,t)|^2),
\label{equi}
\ee which should be close to zero for the inertial modes that exhibit
nearly perfect equipartition.  The summation in \eq{equi} is carried
over all the modes (i.e. $k$'s) that constitute the inertial
range spectrum.  Our simulation results, following the evolution of
\eq{equi}, are shown in Fig 3. We find from our simulations that the
inertial range turbulent fluctuations closely follow the equipartition
that leads to a strong wave activity in the $kd_e < 1$ regime. Despite
the presence of the dispersive whistler waves in this regime, the
inertial range spectrum continues to follow a Kolmogorov-like
$k^{-7/3}$ spectrum.  The equipartition in whistler wave turbulence
has also been reported in our two dimensional work
\citep{dastgeer00a,dastgeer00b,dastgeer05}. Interestingly, our 3D
simulations, describing the equipartition between the velocity and
magnetic field fluctuations, are consistent with the 2D counterpart.
It thus appears that the turbulent equipartition is a robust feature
of whistler waves that is preserved in both 2D and 3D nonlinear mode
coupling interactions.  The spectral cascades of inertial range
turbulent energy is nonetheless determined by the energy cascade per
unit nonlinear time as follows,
\[ \varepsilon \simeq \frac{E}{\tau} \simeq \frac{B_{\ell}^3}{\ell^2}. \]
On assuming that the spectral energy cascade is local in the
wavenumber space \cite{kol,kol,iros}, the energy spectrum per unit
mode yields $ E_k \simeq \varepsilon^{\alpha}k^{\beta}$. On substituting
the energy and energy dissipation rates and equating the powers of
$B_{\ell}$ and $\ell$, we obtain $\alpha = 2/3$ and $\beta = -7/3$. This,
in the $kd_e < 1$ regime, leads to the following expression for the
energy spectrum 
\[ E_k \simeq \varepsilon^{2/3}k^{-7/3}. \]
It can be noted from the dispersion relation, \eq{disp}, that the
group velocity of whistler waves in the $kd_e<1$ regime is $\omega/k_y
\sim \partial \omega/\partial k_y \sim k$ and the dispersion is
$\omega_k \sim \omega_0 k_y k$. Both the quantities are proportional
to the characteristic wavenumber $k$. It is evident from these
relations that the group velocity and dispersion of whistlers are
predominant at the smaller length scales in the $kd_e<1$
regime. Correspondingly, the scaling law for energy cascades is
modified by the short scale spectrum of the whistler waves in the
$kd_e<1$ regime. To determine the effect of small scale whistler waves
on the spectral transfer, we compute the energy transfer rates in the
$kd_e<1$ regime as follows.

The $kd_e<1$ regime comprises the dispersive whistler waves whose
interaction time can be estimated from $\tau_w \simeq \ell/v_g$, where $v_g
$ is the group velocity for the whistler modes. The group velocity of
whistler waves in the $kd_e<1$ regime is $v_g \sim \partial
\omega/\partial k_y \sim k \sim \ell^{-1}$. The interaction time
between two (or more) whistler wave packets thus yields $\tau_w \sim
\ell^2$. The nonlinear energy cascade rates computed as above, i.e.
$\varepsilon \sim E/\tau$, will be modified by the whistler interaction time
as 
\[\varepsilon_w \sim \left(\frac{E}{\tau}\right) \left(\frac{\tau_w}{\tau}\right) \sim \frac{B^4}{\ell^2} .\]
Here $\varepsilon_w$ is the whistler modified energy transfer rates.
On using the Kolmogorov phenomenology that the spectral transfer is
local and depends only on the energy dissipation rates and modes
\cite{kol,kol,iros}, the energy spectrum can be given by $E_k \sim
\varepsilon^\alpha k^\beta$. Upon substituting the energy dissipation
rates, we estimate the spectral energy as $E_k \sim \varepsilon^{1/2}
k^{-2}$. The change in the inertial range spectral slope due to the
whistler waves is referred to as {\it whistler effect}
\cite{biskamp,dastgeer00a,dastgeer00b,dastgeer05}.  By introducing the
whistler time scale in deriving the energy cascade rates
$\varepsilon_w$, it is noteworthy that the spectrum in the $kd_e<1$
regime is modified by the presence of whistler waves and one might infer
that the whistler waves modify the inertial range spectrum from
$k^{-7/3}$ to a more flatter one, i.e. $k^{-2}$.  Although the
difference between the two spectra is small enough to be noticeable in
the 3D simulations (generally because of poor spectral resolutions),
the flattening of the spectrum is not observed in our simulations that
persistently show that the whistler wave spectrum is close to
$k^{-7/3}$. While the spectral resolution in our three dimensional
simulations is not adequate enough to resolve the two distinct
spectra, very high resolution simulations (upto $5120^2$) in 2D
\cite{dastgeer05} suggest that the volume integrated energy spectra
are not affected by the presence of the whistler waves and the
inertial range turbulent fluctuations continue to exhibit a
Kolmogorov-like $k^{-7/3}$ spectrum.  The whistler effect in those
simulations \cite{dastgeer05} is observed to be influential only at
the local region in the inertial range turbulent spectrum. This result
is further consistent with that of MHD turbulence \cite{sheb} where
anisotropy in the spectral space mediated by the Alfv\'en waves (i.e.
the Alfv\'en effect) is explained by virtue of local Fourier mode,
while the volume integrated MHD spectrum exhibits a Kolmogorov-like
\citep{kol} $k^{-5/3}$ power law. The controversy \citep{ iros,
  krai} with regard to the $k^{-5/3}$ or $k^{-3/2}$ MHD spectrum is
not a subject of this paper and we will not discuss it any
further. The reader can however refer to the book by Biskamp
\cite{biskamp03} and the reference therein.

\subsection{$kd_e > 1$: Hydrodynamic-like  regime}
The regime $kd_e > 1$ in whistler wave turbulence corresponds
essentially to a hydrodynamic regime because the EMHD equation,
\eq{emhd3}, in this regime reduces to the Navier Stokes equation that
describes the dynamics of non magnetized hydrodynamic flows. The
energy spectrum in this regime is dominated by the shorter
length-scale turbulent eddies that give rise to the characteristic
spectrum of an incompressible hydrodynamic fluid. The group velocity
of whistlers, in this regime, is small and hence it is expected that
the effect of whistlers will not be present.  For $kd_e > 1$, the
first term in the energy can be neglected and thus
\[ E \sim k^2 B^2 \sim \frac{B_\ell^2}{\ell^2}. \]
where we have used $k \sim 1/\ell$. The energy cascade rates per unit
nonlinear transfer time, $\varepsilon \simeq E/\tau$, in the regime
$kd_e > 1$ lead to $\varepsilon \simeq B_\ell^3/\ell^4$. On using the
Kolmogorov's phenomenology of local spectral cascade, the energy
spectrum of whistler turbulence in the $kd_e > 1$ regime can obtained
as 
\[E_k \sim \varepsilon^{2/3} k^{-5/3},\]
in agreement with our simulations (see Fig 2).  This spectrum is
identical to that of energy in three dimensional incompressible
Navier-Stokes turbulence and further confirms the hydrodynamic nature
of the whistler wave turbulence for the small scale fluctuations in
$kd_e > 1$ regime. While the longer scales ($kd_e<1$ modes) possess
stronger tendency of behaving like whistlers, the shorter scales
($kd_e>1$ modes) act like unmagnetized hydrodynamical eddies where
wave effects are considerably weaker.  Hence whistler wave turbulence
in this regime exhibits the energy spectrum that is essentially
identical to that of hydrodynamic fluid.

\section{Summary}
Three dimensional simulations of turbulent cascades in solar wind
plasma are carried out to quantify the role of whistler waves
corresponding to the inertial range fluctuations that possess
characteristic frequency bigger than the ion gyro frequency ($\omega >
\omega_{ci}$) and length scales smaller than the ion gyro radius
($k\rho_i>1$). In this regime, the solar wind plasma fluctuations
comprise of unmagnetized ions, hence the entire dynamics is governed
by the electron fluid motions. The rotational magnetic field
fluctuations in the presence of a background magnetic field lead to
propagation of dispersive whistler waves in which the wave magnetic
and velocity fields are strongly correlated through the equipartition
($v_e^2 \simeq k^2 B^2$). The latter is employed in our simulations to
quantify the role of whistler waves that are ubiquitously present in
the inertial range in the high frequency ($\omega > \omega_{ci}$)
solar wind plasma. Interestingly we find that despite strong wave
activity in the inertial range, whistler waves do not influence the
inertial range turbulent spectra. Consequently, the turbulent
fluctuations in the inertial range are described by Kolmogorov-like
phenomenology. Thus consistent with the Kolmogorov-like dimensional
argument, we find that turbulent spectra in the $kd_e<1$ and $kd_e>1$
regimes are described respectively by $k^{-7/3}$ and $k^{-5/3}$. Our
results are important particularly in understanding turbulent cascade
corresponding to the high frequency ($\omega > \omega_{ci}$) solar
wind plasma where characteristic fluctuations are comparable to the
electron inertial skin depth.

The support of NASA(NNG-05GH38) and NSF (ATM-0317509) grants is
acknowledged.



\label{lastpage}


\begin{thebibliography}{99}

\bibitem[Alexandrova et al 2007]{alexandrova2007}
Alexandrova, O.; Carbone, V.; Veltri, P.; Sorriso-Valvo, L.
2007, Planet. Space Sci. {\bf 55}, 2224 

\bibitem[Alexandrova et al 2008]{alexandrova2008}
Alexandrova, O.; Carbone, V.; Veltri, P.; Sorriso-Valvo, L. 2008,
Astrophys. J. {\bf 674}, 1153 


\bibitem[Bengt \& Shukla 2008]{Bengt}
Bengt, E., and Shukla, P. K. 2008, AIP Conf. Proc. -- October 15, 2008 -- Volume 1061, pp. 76-83
FRONTIERS IN MODERN PLASMA PHYSICS:  ICTP International Workshop
on the Frontiers of Modern Plasma Physics; DOI:10.1063/1.3013785

\bibitem[Biskamp 2003]{biskamp03}
Biskamp, D. '{\it Magnetohydrodynamic Turbulence}'
Published by Cambridge University Press, 2003

\bibitem[Biskamp et al 1996]{biskamp}
Biskamp, D., Schwarz, E., and 
Drake, J. F. 1996, Phy. Rev. Lett., {\bf 76}  1264

\bibitem[Bhattacharjee et al 1998]{Bhattacharjee1998} 
Bhattacharjee, A., C. S. Ng, and S. R. Spangler, 
{\it Astrophys. J.} {\bf 494}, 409 (1998).

\bibitem[Burman 1975]{burman}
Burman, R. R. 1975,
Astronomical Society of Japan, Publications, 27, 511-513


\bibitem[Cho \& Lazarian 2004]{lazarian}
Cho, Jungyeon; Lazarian, A. 2004, 615, L41


\bibitem[Dastgeer et al. 2000a]{dastgeer00a} 
Dastgeer, S., Das, A., Kaw, P., and Diamond, P. H. 2000a,
	Phys. Plasmas, {\bf 7} 571 

\bibitem[Dastgeer et al. 2000b]{dastgeer00b} 
Dastgeer, S., Das, A., and 
Kaw, P. 2000b, Phys. Plasmas, {\bf 7} 1366 
5

\bibitem[Galtier 2008]{Galtier}
Galtier, S. 2008, J. Geophys. Res., 113, A01102, doi:10.1029/2007JA012821.

\bibitem[Gary et al 2008]{gary}
Gary, S. P., {\it et al.}, 2008, Geophys. Res. Lett. {\bf 35}, L02104

\bibitem[Goldstein et al 1995]{Goldstein1995}
Goldstein, M. L.; Roberts, D. A.; Matthaeus, W. H. 1995, 
Annual Review of Astronomy and Astrophysics, 33, 283


\bibitem[Gottlieb et al 1977]{scheme}
 Gottlieb D., and  Orszag, S. A. 1977, Numerical
Analysis of Spectral Methods, SIAM, Philadelphia

\bibitem[Hasegawa \& Chen 1976]{hasegawa}
Hasegawa, A.,  and Chen, L. 1976, Phys. Rev. Lett. {\bf 36}, 1362



\bibitem[Iroshnikov 1963]{iros}
 Iroshnikov, P. S., {\it Astron. Zh.} {\bf 40}, 742 (1963).


\bibitem[Kingsep et al. 1990]{model}
Kingsep, A. S., Chukbar, K. V.,  and Yankov, V. V. 1990,  in 
Reviews of Plasma Physics (Consultant Bureau, New York) vol 16

\bibitem[Kolmogorov 1941]{kol}  
 Kolmogorov, A. N.  {\it Dokl. Acad. Sci. URSS} {\bf 30}, 301 (1941).


\bibitem[Krafft \& Volokitin 2003]{krafft}	
Krafft, C.; Volokitin, A. 2003,
Annales Geophysicae, vol. 21, Issue 7, pp.1393-1403

\bibitem[Kraichnan 1965]{krai}  
 Kraichnan, R. H., {\it Phys. Fluids} {\bf 8}, 1385 (1965).


\bibitem[Leamon et al 1999]{leamon}
Leamon, R. J.; Ness, N. F.; Smith, C. W.; Wong, H. K., Dynamics of the
Dissipation Range for Solar Wind Magnetic Fluctuations, AIPC, 471,
469, 1999.

\bibitem[Matthaeus \& Brown 1988]{Matthaeus1988} 
 Matthaeus, W. H., and M. Brown, 
{\it Phys Fluids} {\bf 31}, 3634 (1988).




\bibitem[Ng et al 2003]{ng} 
 Ng, C. S., A. Bhattacharjee, K. Germaschewiski,  and S. Galtier, 
{\it Phys. Plasmas} {\bf 10}, 1954 (2003).


\bibitem[Roth 2007]{roth}
Roth, I. 2007, Planetary and Space Science, 55, 2319-2323 


\bibitem[Saito et al 2008]{saito}
Saito, S.,  Gary, S. P.,  Li, H.,  and Narita, Y. 2008, 
 Phys. Plasmas 15, 102305 

\bibitem[Salem et al 2007]{salem}
Salem, C. Bale, S. D., and Maksimovic, M. 2007,
Proc. of the Second Solar Orbiter Workshop, 16-20 October 2006, Athens, Greece
(ESA SP-641, January 2007)




\bibitem[Shaikh \& Zank 2003]{dastgeer03}
Shaikh, D., \& Zank, G. P. 2003, ApJ., 599, 715

\bibitem[Shaikh \& Zank 2005]{dastgeer05}
Shaikh, D., \& Zank, G. P., 2005, Phys. Plasmas, 12, 122310.

\bibitem[Shaikh \& Zank 2006]{dastgeer06}
Shaikh, D., \& Zank, G. P., 2006, ApJ, 640, 195

\bibitem[Shaikh \& Zank 2007]{dastgeer07}
Shaikh, D., \& Zank, G. P., 2007, ApJ, 656, 17

\bibitem[Shaikh \& Shukla 2009]{dastgeer08}
Shaikh, D., \& Shukla, P. K. 2008, 2009, Phys. Rev. Lett., 102, 045004 

\bibitem[Shaikh 2009]{dastgeer09}
Shaikh, D.,  2009, J. Plasma Phys., 75, 117

\bibitem[Shaikh \& Shukla 2008]{dastgeer08a}
Shaikh, D., \& Shukla, P. K. 2008, FRONTIERS IN MODERN PLASMA PHYSICS: 2008 ICTP International Workshop on the Frontiers of Modern Plasma Physics. AIP Conference Proceedings, Volume 1061, pp. 66-75.

\bibitem[Shebalin et al. 1983]{sheb} 
Shebalin, J. V., Matthaeus, W. H.,  and 
  Montgomery, D.  1983, J. Plasma Physics {\bf 3}, 525




\bibitem[Shukla  1978]{Shukla78}
Shukla, P. K. 1978, Nature, 274, 874

\bibitem[Stawicki et al 2005]{Stawicki}
Stawicki, Olaf; Gary, S. Peter; Li, Hui; 2005, JGR, 106, A5,
  8273.


\bibitem[Urrutia et al 2008]{Urrutia}
Urrutia, J. M.,  Stenzel, R. L., and Strohmaier, K. D. 2008, 
 Phys. Plasmas 15, 062109, DOI:10.1063/1.2934680


\bibitem[Vocks et al 2005]{Vocks}
Vocks, C.; Salem, C.; Lin, R. P.; Mann, G. 2005, ApJ,
627, 540


\bibitem[Wei et al 2007]{Wei2007}
Wei, X. H., J. B. Cao, G. C. Zhou, O. Santolık, H. Reme,
I. Dandouras, N. Cornilleau-Wehrlin, E. Lucek, C. M. Carr, and
A. Fazakerley. 2007, J. Geophys. Res., 112, A10225,
doi:10.1029/2006JA011771

\bibitem[Wu \& Yoon 2007]{yoon}
Wu, C. S.,  and  Yoon, P. H. 2007, Phys. Rev. Lett. {\bf 99}, 075001




\end{thebibliography}
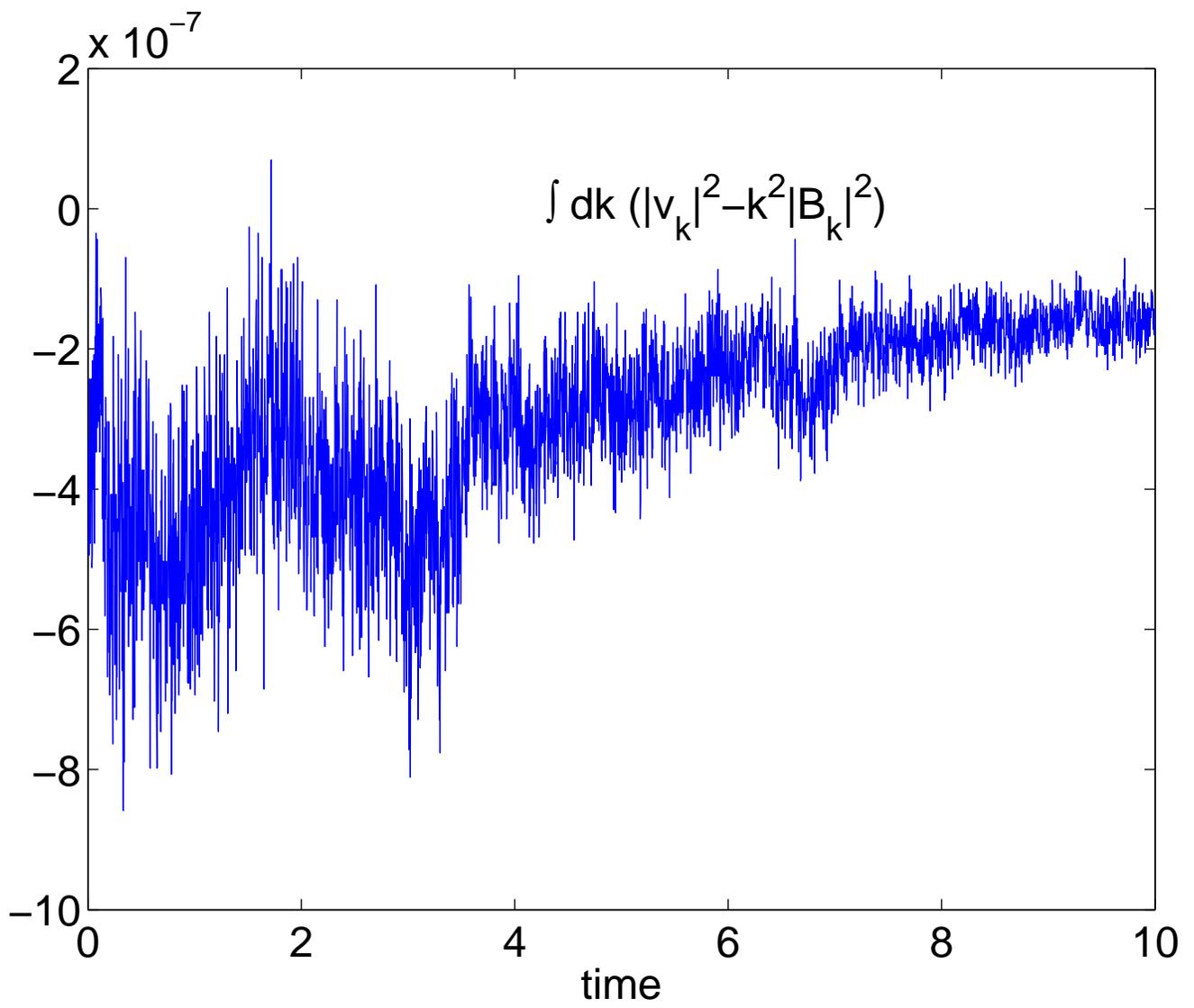
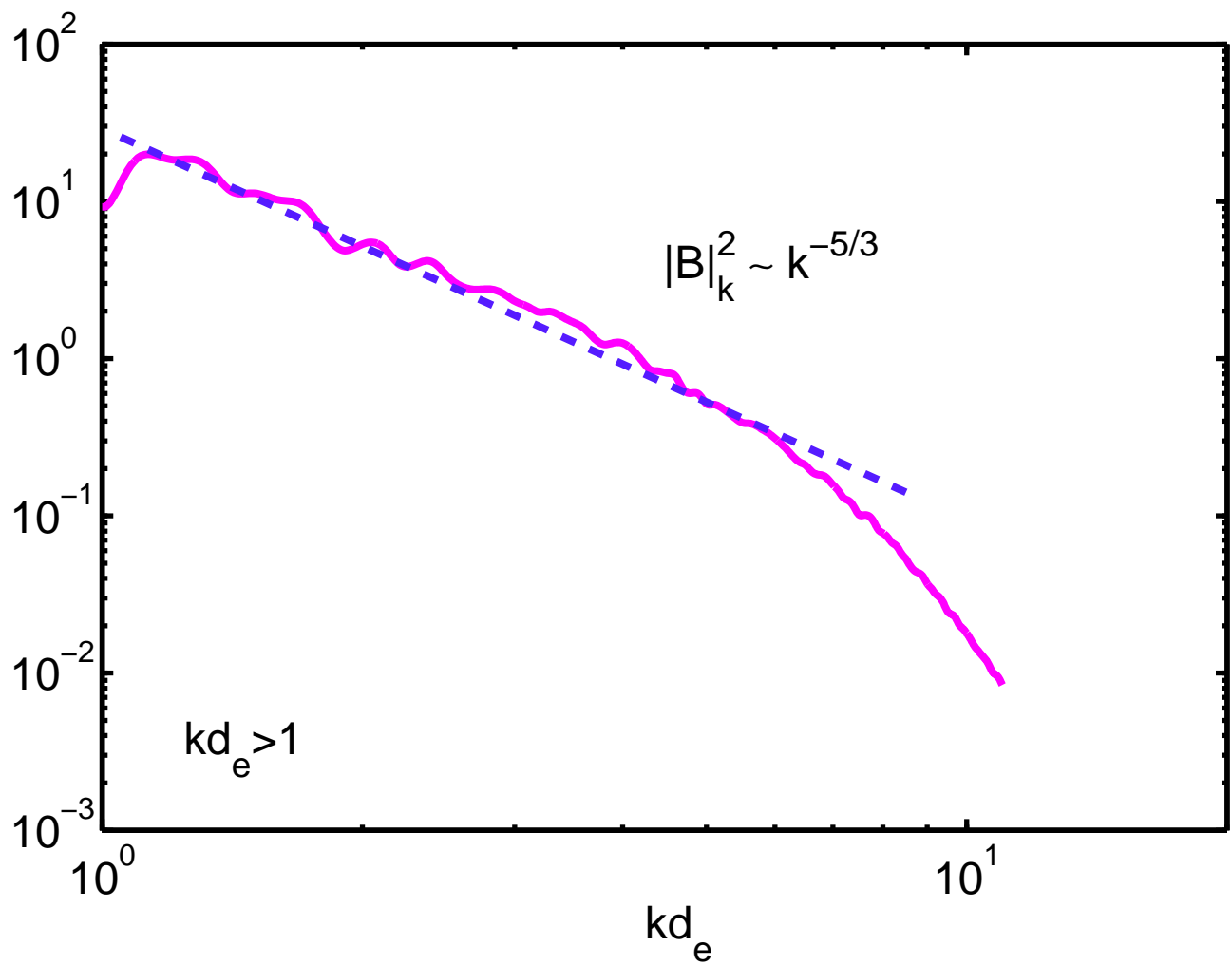
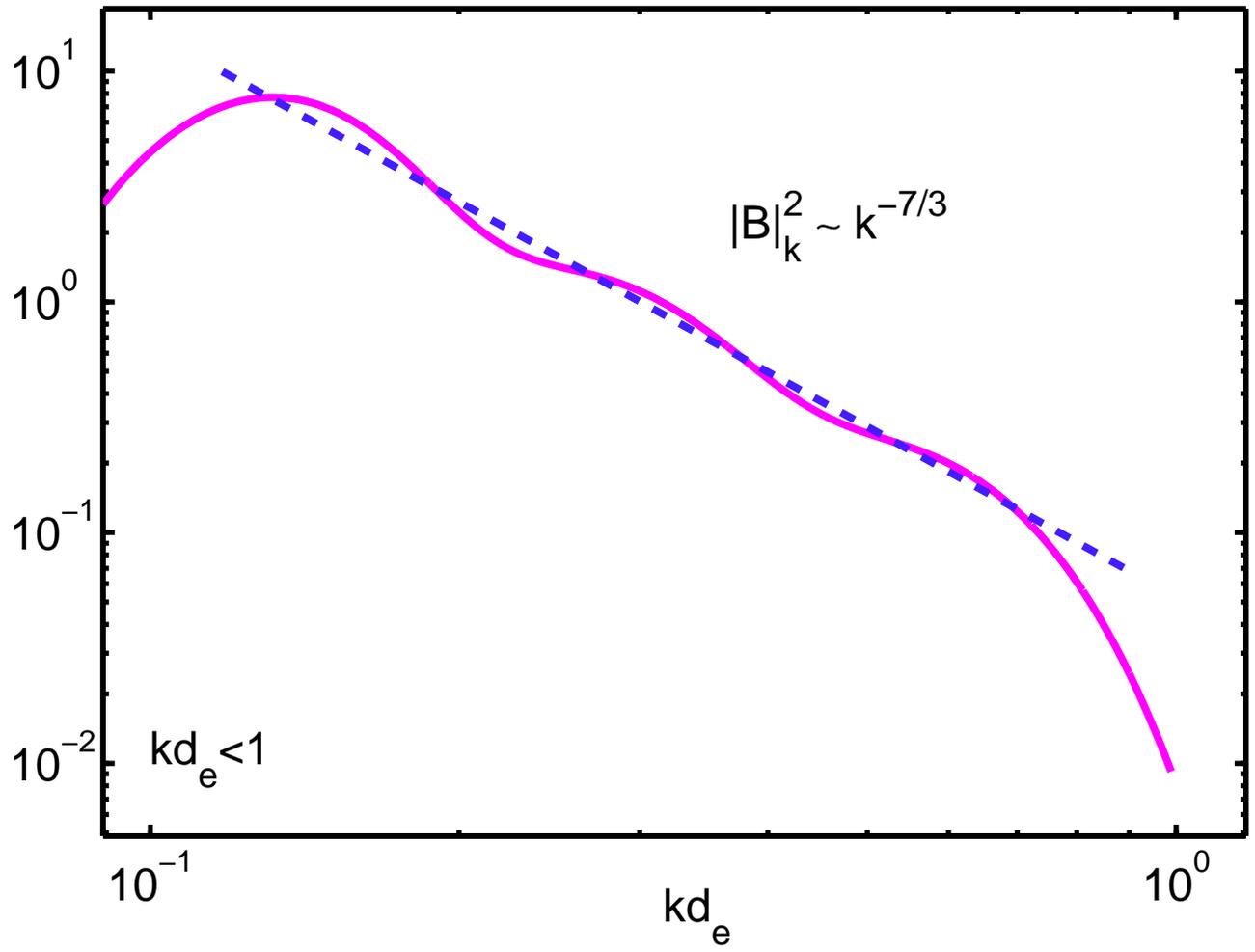
\end{document}